\documentclass[aps,prl,twocolumn,superscriptaddress]{revtex4-2}
\usepackage{graphicx}
\usepackage{amssymb}
\usepackage{amsmath}
\usepackage{multirow}
\usepackage{hyperref}

\begin{document}
\title{Superfluid $\boldsymbol{\beta}$ phase in liquid $^3$He}
\author{V.\,V.\,Dmitriev}
\email{dmitriev@kapitza.ras.ru}
\affiliation{P.L.~Kapitza Institute for Physical Problems of RAS, 119334 Moscow, Russia}
\author{M.\,S.\,Kutuzov}
\affiliation{Metallurg Engineering Ltd., 11415 Tallinn, Estonia}
\author{A.\,A.\,Soldatov}
\affiliation{P.L.~Kapitza Institute for Physical Problems of RAS, 119334 Moscow, Russia}
\author{A.\,N.\,Yudin}
\affiliation{P.L.~Kapitza Institute for Physical Problems of RAS, 119334 Moscow, Russia}

\date{\today}

\begin{abstract}
We report the first observation of superfluid $\beta$ phase of $^3$He. This phase is realized in $^3$He in nematic aerogel in presence of high magnetic field right below the superfluid transition temperature. We use a vibrating aerogel resonator to detect the transition to the $\beta$ phase and measure the region of existence of this phase.
\end{abstract}

\maketitle

{\bf Introduction.---} In bulk superfluid $^3$He (with $p$-wave, spin-triplet Cooper pairing) the free energy and the superfluid transition temperature are degenerate with respect to spin and orbital momentum projections. This allows a variety of superfluid phases with the same transition temperature, but in zero magnetic field only two phases (A and B) with the lowest energy are realized \cite{VW}. Anisotropy of the space may lift the degeneracy and other phases can be stabilized. In particular, the anisotropic scattering of $^3$He quasiparticles may lift the degeneracy with respect to orbital angular momentum projections and make favorable new phases -- polar, polar-distorted A and polar-distorted B phases \cite{AI,S13,fom14,ik15,fom18,vol18,fom20}. These phases were recently observed and investigated in $^3$He confined in nematic aerogel \cite{dmit15,dmit12,dm14,dm16,zhel16,aut16,dmit18,dmit19,dmit20,elt20}. Nematic aerogels consist of nearly parallel strands that results in anisotropic scattering of $^3$He quasiparticles inside the aerogel \cite{asad15,we15}. If the anisotropy is large, then the superfluid transition of $^3$He in nematic aerogel occurs to the polar phase, and, on further cooling, transitions to polar-distorted A and polar-distorted B phases may occur. Both polar and A phases are Equal Spin Pairing (ESP) phases and contain Cooper pairs with only $\pm$1 spin projections on a specific direction ($\uparrow\uparrow$ and $\downarrow\downarrow$ pairs), but, in contrast to the A phase, the polar phase is not chiral and has a Dirac nodal line in the energy spectrum of Bogoliubov quasiparticles in the plane perpendicular to the direction of the aerogel strands.

The degeneracy with respect to spin projections may be lifted by magnetic field. In bulk $^3$He in strong magnetic fields the transition temperature for different spin components is splitted leading to a formation of new (A$_1$ and A$_2$) phases instead of the A phase. Then, instead of the second-order superfluid transition at zero field at $T=T_c$, there are two second-order transitions: to the A$_1$ phase at $T=T_{A1}>T_c$ and to the A$_2$ phase at $T=T_{A2}<T_c$. The $A_1$ phase contains only $\uparrow\uparrow$ pairs and exists in a narrow range of temperatures ($\sim0.02\,T_c$ in field of 10\,kOe) which increases proportionally to the field \cite{A1,osh,isr84,sag84,koj08}. The $A_2$ phase contains also $\downarrow\downarrow$ pairs, which fraction rapidly grows with cooling, and this phase is continuously transformed to the A phase, where fractions of both spin components are equal to each other. Similar splitting should also occur in the polar phase in a strong magnetic field \cite{sur19_1,sur19_2}. On cooling, the superfluid transition should occur to the so-called $\beta$ phase \cite{VW} (or P$_1$ phase in notation of Refs.~\cite{sur19_1,sur19_2}) instead of the pure polar phase. On further cooling, the second-order transition to the distorted $\beta$ (or P$_2$) phase is expected which is continuously transformed to the pure polar phase.

The order parameters of the $\beta$ and distorted $\beta$ phases are
\begin{eqnarray}
A_{\mu j}^{P1}&=&\frac{\Delta_1}{\sqrt{2}}(d_\mu+ie_\mu)m_j, \label{P1} \\
A_{\mu j}^{P2}&=&\frac{\Delta_1}{\sqrt{2}}(d_\mu+ie_\mu)m_j+\frac{\Delta_2}{\sqrt{2}}e^{i\varphi}(d_\mu-ie_\mu)m_j \label{P2}
\end{eqnarray}
correspondingly, where $\Delta_1$ and $\Delta_2$ are gap parameters, $e^{i\varphi}$ is a phase factor, $\bf d$ and $\bf e$ are mutually orthogonal unit vectors in spin space (which are perpendicular to the magnetization), and $\bf m$ is a unit vector in orbital space aligned along the direction of nematic aerogel strands \cite{AI}. From Eqs.~\eqref{P1},\eqref{P2} it follows that orbital parts of order parameters of $\beta$ and distorted $\beta$ phases are the same as in the polar phase, but
the $\beta$ phase contains only $\uparrow\uparrow$ Cooper pairs, while the distorted $\beta$ phase is a condensate of $\uparrow\uparrow$ (the first term in Eq.~(\ref{P2})) and $\downarrow\downarrow$ (the second term in Eq.~(\ref{P2})) pairs. For $\Delta_1=\Delta_2$ Eq.~\eqref{P2} corresponds to the order parameter of the pure polar phase.

Worthy to mark that, although the superfluid A-like phase of $^3$He in silica aerogel corresponds to the A phase of bulk $^3$He,  the A$_1$-A$_2$ splitting in pure $^3$He in silica aerogel was not observed \cite{hal02}. Theory explains this fact by suppression of the splitting due to the presence of solid $^3$He atomic layers on aerogel strands \cite{saul03}.

In this Letter, we present results of high magnetic field experiments in superfluid $^3$He in nematic aerogel, wherein the solid $^3$He layers on the aerogel strands have been replaced by $^4$He. We use a vibrating wire (VW) resonator with the aerogel attached to it, as in previous VW experiments with $^3$He in silica aerogel \cite{bru00,bru01}. We have measured temperature dependencies of resonance properties of the resonator -- the full width at half-maximum (FWHM) and the resonance frequency -- and detected superfluid transitions, which we attribute to transitions between normal and $\beta$, $\beta$ and distorted $\beta$ phases.

{\bf Theory.---} In $^3$He in nematic aerogel, on cooling from the normal phase, a superfluid transition to the $\beta$ phase should occur at the temperature
\begin{equation}\label{tcP1}
T_{P1}=T_{ca}+T_c\eta H,
\end{equation}
where $H$ is the magnetic field,$T_{ca}$ is a superfluid transition temperature of $^3$He in nematic aerogel for $H=0$, and $\eta\sim10^{-3}$\,kOe$^{-1}$ \cite{sur19_1}. On further cooling, the transition to the distorted $\beta$ phase is expected at the temperature
\begin{equation}\label{tcP2}
T_{P2}=T_{ca}-T_c\eta H\frac{\beta_{12345}}{-\beta_{15}},
\end{equation}
where $\beta_{15}=\beta_1+\beta_5$, and so on, $\beta_i$, $i\in\{1,\ldots,5\}$ are coefficients in the Ginzburg-Landau free energy functional \cite{VW}, or beta parameters. On cooling, the distorted $\beta$ phase is continuously transformed to the pure polar phase, that is, $\Delta_2$ in Eq.~\eqref{P2} becomes equal to $\Delta_1$.

From Eqs.~(\ref{tcP1}) and (\ref{tcP2}) we obtain that the temperature range of existence of the $\beta$ phase ($T_{P1}-T_{P2}=T_c\eta H\frac{\beta_{234}}{-\beta_{15}}$) is proportional to $H$, and the P$_1$--P$_2$ splitting is characterized by the following equation:
\begin{equation}\label{P1P2}
\frac{T_{P1}-T_{ca}}{T_{ca}-T_{P2}}=\frac{-\beta_{15}}{\beta_{12345}}.
\end{equation}
Unfortunately, beta parameters of $^3$He in nematic aerogel are unknown. Assuming the bulk $^3$He beta parameters \cite{halpb}, the fraction in Eq.~(\ref{P1P2}) equals 1.36 at 15.4\,bar.

{\bf Methods.---}We used an original sample of mullite nematic aerogel (Metallurg Engineering Ltd.) with density of 150\,mg/cm$^3$, porosity of 95.2\%, and with a size along strands $\approx2.6$\,mm. The strands have a diameter of $\leq14$\,nm (from scanning transmission electron microscope images) and a characteristic separation of 60\,nm. Effective mean free paths of $^3$He quasiparticles in directions parallel and transverse to the strands in the limit of $T=0$ are 900\,nm and 235\,nm correspondingly \cite{dmit20}. The sample for the experiments was cut along the strands from the original sample, so the edges, where the strands begin and end, are not damaged and are perfectly flat: the irregularities are about 100\,nm. Sizes of the obtained rectangular parallelepiped in direction transverse to the strands is $\approx2\times3$\,mm.

\begin{figure}[t]
\centerline{\includegraphics[width=0.5\columnwidth]{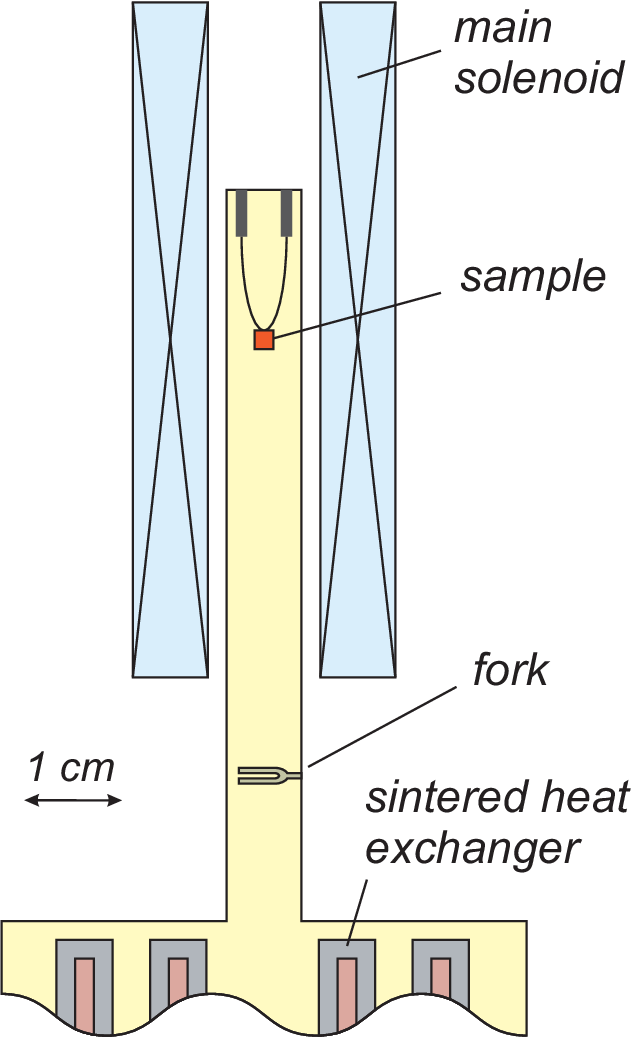}}
\caption{The sketch of the experimental cell.}
\label{ske}
\end{figure}

The experimental sample is glued using a very small amount of Stycast-1266 epoxy resin to 240\,$\mu$m NbTi wire, bent into a shape of an arch with total height of 10\,mm and distance between legs of 4\,mm. Strands of the aerogel were oriented along the oscillatory motion. The aerogel wire is mounted in a cylindrical experimental cell (of internal diameter 6\,mm) made from Stycast-1266 surrounded by a main superconducting solenoid, so that the sample is located at the maximum of the magnetic field (with homogeneity of 0.1\% at distances $\pm3$\,mm). A sketch of the cell is shown in Fig.~\ref{ske}.
\begin{figure}[t]
\centerline{\includegraphics[width=1.0\columnwidth]{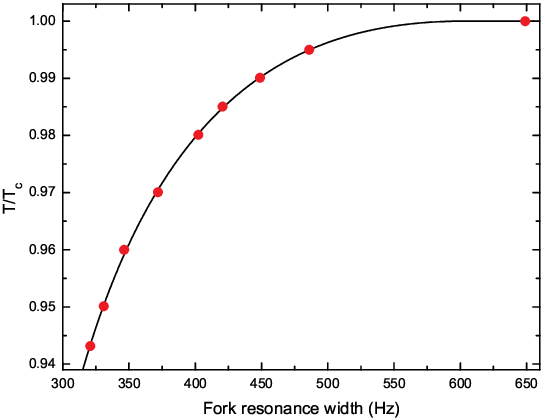}}
\caption{Temperature normalized to $T_c$ versus the fork resonance linewidth in the B phase of superfluid $^3$He. P=15.4\,bar.}
\label{cal}
\end{figure}

In experiments described in the previous version of the manuscript, the temperature was determined under the assumption that on warming, the temperature of $^3$He is changed linearly with time if $T<T_{A2}$. Although a splitting of the $^3$He superfluid transition in nematic aerogel was observed, the splitting asymmetry made us doubt the correctness of this assumption. Therefore, we have replaced the quartz tuning fork in the cell and repeated the experiments. The new fork is larger (originally it was located inside $3\times8$\,mm cylinder) and is installed 12\,mm below the bottom edge of the main solenoid.

The fork resonance linewidth was used to measure the temperature. To calibrate the fork, we used an additional large solenoid (not shown in Fig.~\ref{ske}), which region of homogeneous field is $\approx$100\,mm, so that the fork and the VW are in the same magnetic field. The calibration procedure is the following. We apply the magnetic field generated by the large solenoid in range of 0.3--1.1\,kOe. Then on slow warming we measure the fork resonance width in the B phase just before
B-A transition at $T=T_{BA}$ which is accompanied by jumps in the resonance width and frequency. Subsequent cooling is used to check the transition detection. The B-A and A-B transitions were clearly detected by both the fork and the VW with the time delay less than the times of scanning of both resonances ($\sim$1~minute). The calibration is then obtained using data for dependence of $T_{BA}$ on $H$ given by ``3He calculator'' \cite{3HeC}, which are based on results of Refs.~\cite{PhD,Tang}. The result of the calibration is shown in Fig.~\ref{cal}. In the used temperature range in superfluid $^3$He, the fork was always immersed in the B phase. For this purpose, a relatively small magnetic field of the main solenoid in the fork region was compensated by the large solenoid. At $T>T_c$ the temperature was determined in assumption that the fork resonance linewidth is inversely proportional to $T$ \cite{blaa07}.

The experiments were carried out at a pressure of 15.4\,bar and in magnetic fields 0.5--10.25\,kOe generated by the main solenoid. In order to avoid solid $^3$He atomic layers on the aerogel strands and to stabilize the polar phase in low fields \cite{dmit18}, we had added 1.55\,mmole of $^4$He into the empty cell at $T\leq100$\,mK and then filled it with $^3$He. This amount of $^4$He is enough to completely remove solid $^3$He from aerogel strands and, according to our estimations, corresponds to 2.5--3.2 atomic layers of $^4$He coverage \cite{we20}.

The necessary temperatures were obtained by a nuclear demagnetization cryostat. A measurement procedure of the aerogel resonator is similar to that of a conventional wire resonator \cite{CHH}. An alternating current, with amplitude varying from 0.4 to 8.9\,mA (depending on $H$ and being set to compensate influence of $H$ on the amplitude of oscillations), is passed through the VW. The Lorentz force sets the wire into oscillations. The resonance frequency of our VW resonator in vacuum is 621\,Hz. Motions of the VW in the magnetic field generates a Faraday voltage. This voltage was amplified by a room-temperature step-up transformer {1:30} and measured with a lock-in amplifier. In-phase and quadrature signals (obtained by sweeping the frequency of the driving current) were joint fitted to Lorentz curves in order to extract the FWHM and the resonance frequency. In liquid $^3$He the maximum velocity of the WV in the used range of temperatures did not exceed 0.2\,mm/s. In a given field additional experiments with 2 times smaller excitation current were also done and showed the same results.

Similar mullite samples (cutted from the same original nematic aerogel sample) had been used in nuclear magnetic resonance experiments in $^3$He \cite{dmit19} and in VW experiments in low magnetic fields \cite{dmitVW}. Correspondingly, we can expect that the present sample should have nearly the same $T_{ca}$ ($\approx 0.985\,T_c$ at 15.4\,bar) and the temperature width of the superfluid transition about 0.002\,$T_{ca}$. We note that in experiments described in Ref.~\cite{dmitVW} an additional (the second) resonance mode had been observed, existing only below $T_{ca}$. This second mode is an analog of the second-sound-like mode (called also as slow sound mode) observed in silica aerogel in superfluid helium \cite{McK,gol99} and corresponds to motions in opposite directions of the superfluid component inside the aerogel and the normal component (together with the aerogel strands). On cooling from $T=T_{ca}$, the resonant frequency of this additional mode very rapidly increases from 0 up to $\sim1.6$\,kHz, and in a narrow temperature range below (but very close to) $T_{ca}$ becomes close to the resonance frequency of the main mechanical VW resonance resulting in an interaction of these modes (see Ref.~\cite{dmitVW} for details). In present experiments we focused on measurements of the main resonance, which intensity is significantly greater.

\begin{figure}[t]
\centerline{\includegraphics[width=1.0\columnwidth]{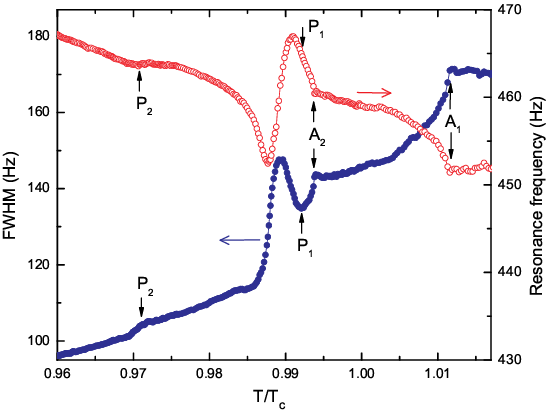}}
\caption{Temperature dependencies of FWHM (filled circles) and frequency (open circles) of the main resonance of the VW resonator measured in magnetic field of 10.25\,kOe at excitation current of 0.4\,mA. Arrows indicate the features we associate with $T_{P2}$, $T_{P1}$, $T_{A2}$, and $T_{A1}$. $T_c$ is a superfluid transition temperature of bulk $^3$He in zero magnetic field. P=15.4\,bar.}
\label{WFr}
\end{figure}

{\bf Results.---}Most of the experiments were done on a slow (0.002--0.004\,$T_c$ per hour) warming of the cell.
In Fig.~\ref{WFr} we show results obtained in magnetic field of 10.25\,kOe, where we measured the FWHM and the frequency of the main resonance of the VW. In Fig.~\ref{WFr} we mark features (A$_1$, A$_2$, P$_1$, and P$_2$) which we ascribe to superfluid transitions at temperatures $T_{A1}$, $T_{A2}$, $T_{P1}$, and $T_{P2}$.
\begin{figure}[t]
\centerline{\includegraphics[width=1.0\columnwidth]{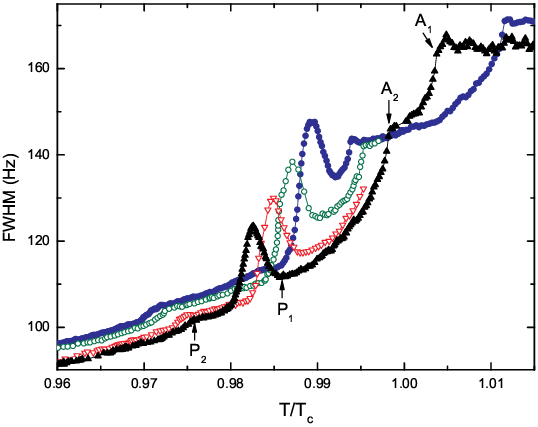}}
\caption{Temperature dependencies of the FWHM of the main resonance of the VW resonator measured in magnetic fields of 10.25\,kOe (filled circles), 8.2\,kOe (open circles), 6.15\,kOe (open triangles), and 4.1\,kOe (filled triangles) at corresponding excitation currents of 0.4\,mA, 0.5\,mA, 0.67\,mA, and 1\,mA.
For a better view, the arrows mark P$_1$, P$_2$, A$_1$, and A$_2$ features only for $H$=4.1\,kOe. P=15.4\,bar.}
\label{4field}
\end{figure}
\begin{figure}[t]
\centerline{\includegraphics[width=1.0\columnwidth]{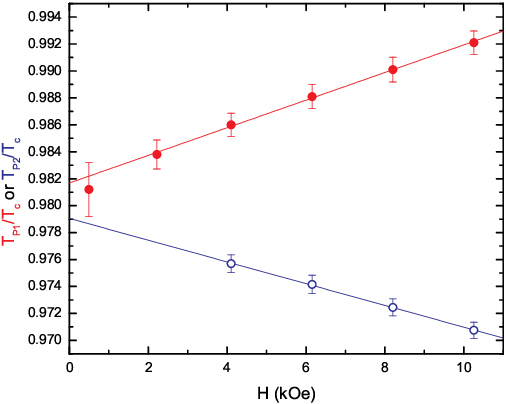}}
\caption{P$_1$--P$_2$ splitting of the superfluid transition of $^3$He in nematic aerogel in magnetic field. Open and filled circles indicate transitions between distorted $\beta$ and $\beta$, $\beta$ and normal phases respectively. Lines are linear approximations of experimental data.}
\label{split}
\end{figure}

Let us consider these features with decreasing temperature.
At $T>T_{A1}$ both bulk $^3$He and $^3$He in aerogel are in the normal state: on cooling, the FWHM slowly increases and the frequency is slowly decreases. The superfluid transition to the A$_1$ phase in bulk $^3$He occurs at $T=T_{A1}$. The accuracy of determination of $T_{A1}$ is only $\pm$0.003\,$T_c$ because at $T=T_{A1}$ the fork is in the normal phase where its resonance linewidth changes very slowly. Below $T_{A1}$ the FWHM decreases and at $T=T_{A2}$ the transition to the A$_2$ phase occurs. According to our temperature calibration, $T_{A2}$ is slightly higher (by 0.0015\,$T_c$) than it follows from Ref.~\cite{isr84}. On further cooling, the FWHM decreases more rapidly but below $T=T_{P1}$ it starts to increase that can be due to only the superfluid transition of $^3$He in aerogel. In the given magnetic field this transition should be to the $\beta$ phase. At lower temperature (at $T=T_{P2}$) we observe ``step'' on the FWHM plot or ``kink'' on the resonance frequency plot, which we refer to the transition between the $\beta$ phase and the distorted $\beta$ phase existing at $T<T_{P2}$. We note that on cooling below $T=T_{P2}$ the intensity of the second resonance mode starts to rapidly grow, but in the region of existence of the $\beta$ phase ($T_{P2}<T<T_{P1}$) its intensity is very small. We assume that in the $\beta$ phase this mode is less excited and, in comparison with experiments described in Ref.~\cite{dmitVW}, we did not observe a clear repulsion between the main and the second resonance modes at $T\approx T_{P1}$, that is near the superfluid transition of $^3$He in aerogel. However, the interaction between these modes in the $\beta$ phase remains, and just below $T=T_{P1}$ in the main resonance we observe a peak-like change of the linewidth as well as the rapid change of the resonance frequency.

In Fig.~\ref{4field} we show temperature dependencies of the FWHM of the main VW resonance obtained in different magnetic fields. As it was expected, the temperature range of existence of the $\beta$ phase ($T_{P1}-T_{P2}$) is decreased in lower magnetic field.

In Fig.~\ref{split} we summarize results of our experiments and show the measured at 15.4\,bar dependencies of $T_{P1}$, $T_{P2}$ on the applied magnetic field. The results are well fitted by linear functions as it follows from the theory. The ratio of slopes of the fit lines ($(dT_{P1}/dH)/(-dT_{P2}/dH)$) equals 1.27. From Eq.~\eqref{P1P2} this ratio is expected to be equal to 1.36 if we consider the beta parameters of bulk $^3$He \cite{halpb}. We note that the linear fits do not match at $H=0$. This discrepancy may be due to a final width ($\sim0.002\,T_c$) of the superfluid transition of $^3$He in aerogel. It may result in a systematic error in determination of $T_{P1}$ of the same order. In any case, it can be seen that the temperature range of existence of the $\beta$ phase is nearly proportional to $H$ and the value of the splitting is of the same order as it was observed in bulk A phase \cite{isr84,sag84}.

{\bf Conclusions.---}Using the VW techniques, we have observed the $\beta$ phase and measured the P$_1$--P$_2$ splitting of the superfluid transition temperature of $^3$He in nematic aerogel in strong magnetic fields. We have found that the temperature range of existence of the $\beta$ phase is nearly proportional to $H$.

This work was supported by the Russian Science Foundation (project no. 18-12-00384). We are grateful to I.A.~Fomin and E.V.~Surovtsev for useful discussions.

\end{document}